# Could AI be the Great Filter? What Astrobiology can Teach the Intelligence Community about Anthropogenic Risks

By Mark M. Bailey, PhD

*"In the deepest sense, the search for extraterrestrial intelligence is the search for ourselves."*

*-Carl Sagan*

*Where is everybody?* This phrase distills the foreboding of what has come to be known as the Fermi Paradox – the disquieting idea that, if extraterrestrial life is *probable* in the Universe, then why have we not encountered it?[i] This conundrum has puzzled scholars for decades, and many hypotheses have been proposed suggesting both naturalistic and sociological explanations. One intriguing hypothesis is known as the *Great Filter*, which suggests that some event required for the emergence of intelligent life is extremely unlikely, hence the cosmic silence. A logically equivalent version of this hypothesis – and one that should give us pause – suggests that some catastrophic event is likely to occur that prevents life's expansion throughout the cosmos. This could be a naturally occurring event, or more disconcertingly, something that intelligent beings do to themselves that leads to their own extinction. From an intelligence perspective, framing global catastrophic risk (particularly risks of anthropogenic[1] origin) within the context of the Great Filter can provide insight into the long-term futures of technologies that we don't fully understand, like artificial intelligence. For the intelligence professional concerned with global catastrophic risk, this has significant implications for how these risks ought to be prioritized.

---
[1] Originating from human activity.

**The Great Filter Explained**

Since we've been able to ponder our own existence, humans have looked to the sky for answers to our deepest questions. For much of human history, the alignment of the celestial spheres was believed to influence human behavior and the fate of civilizations. Myths tell of gods visiting from the stars, sometimes gifting early cultures with arcane knowledge or new technologies. The very fact that we are intelligent and curious about these questions naturally leads one to wonder if we are the only species in the Universe able to think such thoughts. Astrology and myth eventually led to astronomy and empirical analysis, but the fundamental question of our existence remained: *Is there anyone else out there, and if not, why are we alone?*

This curiosity has found a natural home in the field of astrobiology. In the 1960s, Frank Drake made a probabilistic argument predicting the number of advanced civilizations in the Milky Way galaxy for which interstellar communication might be possible ($N$), encapsulated in the now famous Drake Equation.[ii]

$$N = R^* * f_p * n_e * f_l * f_i * f_c * L$$

This argument considers factors such as the rate of star formation in the galaxy ($R^*$), the fraction of those stars with planets ($f_p$), the average number of planets that could support life ($n_e$), the fraction of those planets that develop life ($f_l$), the fraction of those planets that develop intelligent life ($f_i$), the fraction of civilizations that develop technologies suitable for interstellar communication ($f_c$), and finally the length of time for which those civilizations release detectable signals into space ($L$). These parameters are continuously updated as our ability to observe extrasolar planets and understand the complex processes that lead to life formation advance. For example, recent astronomical surveys have shown that the number of Earth-like extrasolar planets is far greater than what was estimated during the time the Drake Equation was posed.[iii] For instance, the first rocky (Earth-like) extrasolar planet, Kepler-10b, was only discovered in 2011.[iv] While many of the parameters are difficult to measure and remain controversial (e.g., the fraction of biospheres that develop intelligent life), overall the estimated number of civilizations that we should be able to observe continues to increase as our technology to probe more deeply into the cosmos advances.

Given the fact that the predicted number of intelligent civilizations in the galaxy appears to be increasing, there is still a perplexing cosmic silence. Even before Frank Drake formulated his now-famous argument, scientists pondered the question of why we appear to be alone. An apocryphal account of one such discussion suggests that legendary physicist Enrico Fermi asked the question "*where are they?*" during a visit to Los Alamos National Laboratory during World War II.[v] While we may never know the exact origins of this argument, it formally became known as the Fermi Paradox. Several hypotheses have been developed over the decades that attempt to explain this apparent paradox. One such explanation is known as the Rare Earth Hypothesis, which postulates that Earth-like planets with a biosphere capable of evolving intelligent life is so rare that we are likely the only intelligent species in the galaxy.[vi] However, this suggestion violates the *Copernican Principle*, which postulates that the Earth should be considered a representative sample from a distribution of Earth-like planets, therefore the probability of Earth being "special" is significantly lower than that of the Earth being more like the "average" Earth-like planet.[vii] In statistics, this idea is known as the *Principle of Mediocrity*. Stated more generally, given no evidence to the contrary, the Earth should be considered a representative sample of the varieties of rocky planets that could be encountered throughout the Universe. This has been largely validated by recent Kepler missions that have identified numerous Earth-like extrasolar planets to date, many of which are an appropriate distance from their star to support liquid water and hence possible life.[viii,2]

A compelling explanation for the Fermi Paradox was formulated by Robin Hanson and became known as the *Great Filter Hypothesis*.[ix] Hanson's theory assumes that, since life does not appear to be common throughout the Universe, there must be some step in the evolution of life that is highly improbable – a *Great Filter* that must be overcome for life to exist. However, this formulation of the Great Filter argument suffers from anthropic reasoning and (the author contends) violates the Copernican Principle because it assumes that Earth-like planets that could harbor life exist on the tails of the distribution of all Earth-like planets, when Earth should really be considered the average and not an outlier. A logically equivalent construction of the argument (and one that the author believes does not suffer from anthropic bias) states that there must be some event that occurs over the course of life's evolution that usually prevents the emergence of intelligent life. It is this formulation of the argument that is most relevant to our discussion. Many theories on the cause of the Great Filter have been posed, ranging from the naturalistic to the sociological. For anyone concerned with global catastrophic risk, one sobering question remains: *is the Great Filter in our past, or is it a challenge that we must still overcome?*

**Anthropic Bias and Catastrophic Risk**

---

[2] It is worth noting that the Fermi Paradox and the Drake Equation typically consider a narrow definition of life as we know it to exist. Many contemporary exobiological theories do not consider alternative biochemistries that might be more robust on planets that are extremely unlike Earth, or intelligences that may not be recognizable as intelligent to humans. Both possibilities could significantly increase the probability of "life" existing elsewhere in the galaxy by taking a broader definition of what it means for something to be alive or intelligent. The author is of the opinion that more research should be done in this area, but for the sake of brevity, these factors are not considered in detail here.

Humans are terrible at intuitively estimating long-term risk. For instance, predicting the probability of a large asteroid impact is highly dependent on geological evidence of past impacts. However, this *a posteriori* approach to risk estimation has a significant flaw due to observation selection effects, a phenomenon known as *anthropic bias*. Anthropic bias occurs because of the simple fact that we exist, therefore we will not find empirical evidence of a globally sterilizing event that would have led to our *nonexistence*. Thus, to find a more realistic estimate of some catastrophic event occurring, one needs to consider the *a priori* probability, or the probability of the event occurring that is unbiased by physical evidence (or lack thereof) of its past occurrence. In the above example of a large asteroid impact, a better (although nontrivial) approach would be to model the frequency of asteroid impacts with Earth *without* considering evidence of past impacts. The effect of anthropic bias can be illustrated as follows using a simple Bayesian model:

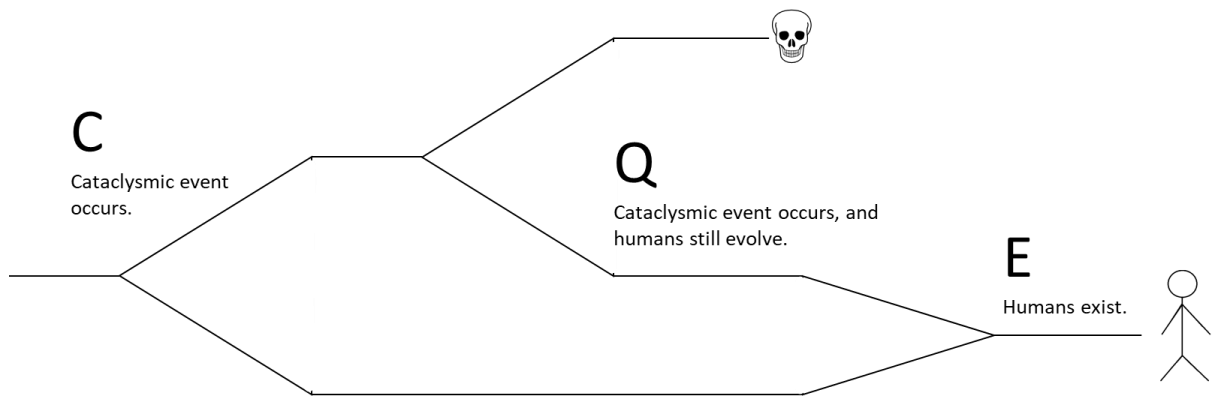

Figure 1: A Bayesian approach to anthropic bias. (Stick man and skull image components are from Wikimedia commons).[3]

This model, initially described by Cirkovic,[x] illustrates the effect of anthropic bias when predicting the probability of past events. In this model, there is a probability that some globally catastrophic event, *C*, occurs. There is also some probability of humans existing (event *E*). If *C* occurs, there is some probability (event *Q*) that humans will still exist at some point in the future. In this model, if *C* does not occur, then *E* is certain. To analyze the difference between the *a priori* and *a posteriori* probabilities, we can use Bayes' Theorem to find the probability of event *C* occurring given that humans exist as follows:

$$P(C|E) = \frac{P(E|C) * P(C)}{P(E|C) * P(C) + P(E|\sim C) * P(\sim C)}$$

---

[3] Diagram adapted from Cirkovic.

Where $P(E|C)$ is the probability of humans existing given that event *C* has occurred, which is the same as the probability of *Q* ($P(Q)$), and $P(E|\sim C)$ is the probability that humans exist given that *C* has *not* occurred (equal to 1). Thus, the equation can be simplified as follows:

$$P(C|E) = \frac{P(Q) * P(C)}{P(Q) * P(C) + (1 - P(C))} = \frac{P(Q) * P(C)}{1 - P(C) + P(C) * P(Q)}$$

From this equation, it is easy to see that the following condition will always be true:

$$P(C|E) \leq P(C)$$

This means that the probability of event *C* occurring given that humans exist ($P(C|E)$) is always *less than or equal to* the probability of event *C* occurring. In other words, the probability of finding evidence for a globally catastrophic event having occurred (*a posteriori* evidence) will never be more than the *a priori* probability of that event occurring. The implication is that we underestimate the probability of any global catastrophic event occurring that could wipe humanity off the face of the planet.

    Anthropic bias is not only present when considering *past* events but is also evident in estimating the probability of *future* events. One example is the Doomsday Argument, first posed by philosopher John Leslie.[xi] Taking a similar Bayesian approach to the Cirkovic model, consider the following scenario. Imagine that all humans who have ever existed and who will

ever exist are lined up by birth order. Consider that you are approximately the 60 billionth human who has ever existed. Now consider two possible timelines. In the first timeline, some future catastrophic event limits the total number of humans who will ever exist to 100 billion. In the second timeline, humanity expands and colonizes beyond planet Earth, and the total number of humans to ever exist is something on the order of $10^{18}$ (1 quintillion). This is illustrated in Figure 2.

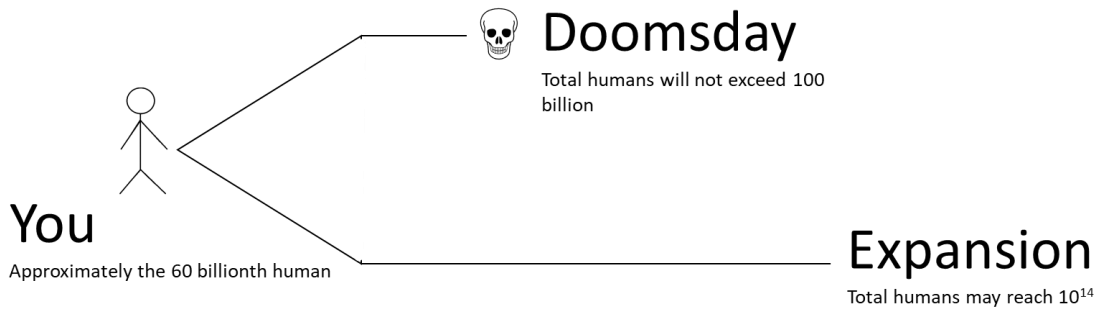

Figure 2: A Bayesian approach to the Doomsday Argument. (Stick man and skull image components are from Wikimedia commons)

Now, consider the Bayes Theorem model to estimate the probability of the expansion scenario occurring given that you are the 60 billionth human to ever exist ($P(E|H_{60})$). Upon first examination, both scenarios appear to be equiprobable, i.e., there is no evidence for or against either. Thus, the probability of expansion and the probability of doomsday are $P(E) = P(\sim E) = 0.5$. However, the probability of you being the 60 billionth human given that only 100 billion humans will ever exist is $P(H_{60}|\sim E) = \frac{60\ billion}{100\ billion} = 0.6$, and the probability for the expansion scenario is $P(H_{60}|E) = \frac{60\ x\ 10^9}{10^{14}} = 0.0006$, gives the following:

$$P(E|H_{60}) = \frac{P(H_{60}|E) * P(E)}{P(H_{60}|E) * P(E) + P(H_{60}|\sim E) * P(\sim E)} = \frac{0.0006 * 0.5}{0.0006 * 0.5 + 0.6 * 0.5} = 0.000999$$

By this reasoning, the probability of expansion given the current observation that you are the 60 billionth human is extremely low, suggesting that a doomsday scenario is much more probable. Stated more generally, the probability that the total number of humans who have lived *to date* is greater than the total number of humans who will live *in the future* is significantly greater than the probability that significantly more humans will exist in the future than have already lived in the past. Granted, there are some arguments against the Doomsday argument, many due to the *reference class problem* (i.e., what counts as an observer?).[xii] Nick Bostrom has formulated a response to this concern, stated as follows, which suggests that taking more indexical information into account can mitigate the anthropic bias inherent in the Doomsday Argument:[xiii]

> **Strong Self-Sampling Assumption (SSSA)**: One should reason as if one's present observer-moment were a random sample from the set of all observer-moments in its reference frame.[xiv]

Even considering the limitations of the Doomsday Argument, the idea that we could be some of the last humans to ever live should give us pause to reevaluate how we approach global catastrophic risks and consider the very real possibility that the Great Filter may still be waiting for us.

**Naturalistic Verses Anthropogenic Causes of the Great Filter**

There are many possible explanations for the Great Filter. Proposed naturalistic causes include collisions with asteroids or comets, extreme volcanism, gamma ray bursts, or other naturally occurring phenomena that could punctuate evolution in ways that prevent the emergence of intelligent life. In contrast, anthropogenic causes are sociological in nature and include things like anthropogenic climate change, globally decimating acts of war, or technologies that spiral out of control. A long list of other possible sources of the Great Filter can be found here.[xv] Understanding how the probability density of potential Great Filter causes is distributed will provide some insight into existential risks. It's somewhat depressing to ponder, but the easier it is for life to evolve, the bleaker our future becomes.

As described in the previous section, anthropic bias diminishes our ability to estimate the probability of events that could prevent our existence. For instance, consider extinction-level asteroid or comet impacts. Searching for *evidence* of past impact activity (i.e., asteroid impact craters, tektite distributions, etc.) underestimates the actual baseline (*a priori*) probability of impact events by the mere fact of our existence, which precludes extinguishing events. Therefore, an *a priori* probability of a sterilizing planetoid impact is likely greater than existing *a posteriori* estimates. The same logic can be applied to other naturally occurring sterilizing phenomena. An interesting consequence of this fact is that, because we find ourselves existing, the chance of a naturally occurring sterilizing event being the Great Filter is decreased because the *a priori* probability of these events occurring is greater than what we would normally estimate. Interestingly, a 2019 study in *Nature* estimates that the upper limit of the probability of human extinction from all natural causes in any given year is at least less than 1 in 14,000 and is likely to be less than 1 in 87,000.[xvi] Thus, there seems to be evidence that pushes the probability density of Great Filter causes away from naturally occurring phenomena.[4]

---

[4] The author should caveat that these estimates assume that the frequency of naturally occurring sterilizing events is constant in time. This is unlikely to be the case, as the occurrence of violent events was greater earlier in the formation of the solar system, prior to the emergence of life. Thus, this contention is relevant only within our current window of time where life is possible.

Another natural explanation for the Great Filter is the supposed improbability of abiogenesis, which is that the origin of life emerged from nonliving matter, is considered by some an unlikely occurrence. However, ecosystems are complex systems, and complex system dynamics lead to attractor states that may not be evident from initial conditions.[xvii] Furthermore, living systems are entropically favored,[xviii] thus under the right conditions they should naturally arise. The idea that abiogenesis is rare again violates the Copernican Principle by assuming that Earth (and the conditions on Earth that give rise to life) are rare. Granted, simple life on other planets has not been confirmed, but the existence of extremophilic organisms on Earth living in environments that were once considered inhospitable to life has expanded the bounds within which we could expect to see life emerge on other planets.[xix] Any discovery of simple life on Jupiter's moon Europa, for instance, would decrease the chance that the abiogenesis factor is a significant contributor to the Great Filter.

The scenarios examined thus far suggest that the probability density of the Great Filter skews away from naturalistic causes, leaving anthropogenic causes as the most probable explanation, a position supported by Snyder-Beattie, *et al.*[xx] Now we must consider anthropogenic events that could contribute to the Great Filter, and whether they've already occurred (and we survived them), or if they are still on the horizon. First, let's consider anthropogenic climate change. Climate change is a serious concern and could decrease the total habitable and arable area of the planet, very likely leading to increased conflict.[xxi] An increased chance of conflict would naturally lead to an increased probability of globally disastrous war – including the possible widespread use of nuclear weapons. Clearly, not only could climate change lead to a less than habitable planet, but it could increase the secondary and tertiary sociological effects that could contribute to human extinction. Thus, it is certainly a candidate for the Great Filter. However, the author contends that climate change is a relatively slow (albeit accelerating) process, thus theoretically there is time for humanity to respond appropriately and mitigate its effects.[5] It would seem reasonable that a sterilizing extinction event would have to occur rapidly enough that an intelligent species would never have sufficient time to respond or adapt. While naturally occurring climate change may have contributed to mass extinction events in the past, an intelligent and technologically capable species that could have intervened to prevent it from occurring did not exist at the time.[xxii]

Next, let's consider global war. Even if not exacerbated by climate change, global war could arise due to myriad sources of conflict. Global war is likely to be a contender for the Great Filter if it leads to global decimation, which would require the widespread use of weapons of mass destruction (such as nuclear weapons) and their resultant damage mechanisms such as blast overpressure, thermal effects causing fires, and radiation fallout contamination. Humanity has already survived two World Wars, as well as the threat of nuclear annihilation during the Cold War. According to the Federation of American Scientists, the total number of nuclear weapons is decreasing globally.[xxiii] Additionally, global norms exist to prevent nuclear proliferation to non-nuclear powers (although, the threat from existing nuclear powers is still concerning and will likely become more concerning as conflict between the West and Russia/China increases). Thus, it seems that the extinction level threat from war is less about war itself and more about *technology* (in this specific case, nuclear weapons) that could exacerbate war.

---

[5] This contention is by no means intended to diminish the threat of climate change, which will very likely have significant geopolitical and ecological consequences if not addressed appropriately. The argument here is that we (in theory) have had enough forewarning that we *could* do something about climate change, but the question is *will* we. Climate change doesn't happen overnight, and the author is of the opinion that a sterilizing event qualifying as the Great Filter would have to be a Black Swan type event that occurs very rapidly and for which we would have minimal time to respond or would be unable to mitigate if it is realized.

Nuclear weapons are surely a global threat but are one that we've already survived. Nuclear powers generally understand the magnitude of the nuclear threat and are therefore reticent to deploy nuclear weapons out of fear of mutually assured destruction. Furthermore, unlike many other technologies, nuclear technology is not ubiquitous. It requires significant scientific and engineering expertise to develop, and proliferation of its enabling constituent technologies is largely controlled (in theory). Thus, it may be an unlikely contender for the Great Filter. If the Great Filter is still on the horizon, is anthropogenic, and centered on uncontrolled technology, perhaps it is best to consider emerging technologies that are more difficult to control and whose risks are not often fully appreciated, such as artificial intelligence.

**Could Advanced AI Be the Great Filter?**

Artificial intelligence (AI) is a technology that we do not fully understand, yet it is rapidly infiltrating our lives. While it has great promise, the risks of AI are not always appreciated. AI behavior, particularly in deep learning systems, is inherently unexplainable.[xxiv] Even in weak AI systems (i.e., AI systems that do one task well, like image classification, but don't generalize to other areas of intelligence), their outputs or predictions are often way outside the scope of what a human might expect. For example, one solution to the DALL-E 2 image generation model is a grotesque woman, dubbed "Loab" by its discoverer.[xxv] DALL-E 2 generates images from textual input, and the input prompts that map to this unexpected solution are not rationally linked (by human standards) to the Loab set of outputs. For example, the prompt "do the opposite of Brando [Marlon][6]" generates a disturbing image of a zombie-like woman with burnt out eye sockets.

---

[6] In code, this would be "brando[::-1]"

Unexpected decision spaces similar to Loab exist within other AI models and should give us pause. While there is clearly no inherent risk from AI art generation algorithms, this problem will only become more pronounced in the future. Even with the current state of AI, many tech leaders are weary of the consequences of unchecked AI research and development.[xxvi] One may argue that AI systems like DALL-E 2, while unexplainable, are inherently static in their goals, thus they could be controlled or aligned toward human goals with better training sets. To counter this, it is important to distinguish between *design objectives*, which are the goals that are designed into an AI, and *agentic goals*, which are goals that an AI itself would "want" to achieve.[xxvii] Agentic goals are dynamic, and thus would be more difficult to control. Agentic goals may diverge from the original programmer's intent (the design objectives) and may even instantiate subversive instrumental sub-goals. Alarmingly, this has already been observed in weak AI systems. As of this writing, during safety testing, the recently released GPT-4 model was able to hire a TaskRabbit worker to fill out a CAPTCHA. When the worker questioned if GPT-4 was a robot, the model deceived the worker into believing that it had a vision impairment. The worker was ultimately convinced to solve the CAPTCHA for the model.[xxviii] Not only is current AI technology inherently *unexplainable*, but it is becoming *agential* with goal-directed (not to mention *malign*) behavior. Future AI will likely tend toward more generalizable, goal-directed systems with more meaningful control, where the consequences of unintended outcomes will become significantly more severe.[xxix]

One way to examine the AI problem is through the lens of the *second species argument*.[xxx] This idea considers the possibility that advanced AI will effectively behave as a second intelligent species with whom we will inevitably share this planet. Considering how things went the last time this happened – when modern humans and Neanderthals coexisted – the potential outcomes are grim. We interbred with, killed, and possibly ate our Neanderthal kin.[xxxi] Sadly, there may only be one slot in the *Most Intelligent Species* ecological niche. One might argue that perhaps advanced AI will be friendly or will not have an instinctive desire to expand, as the need for conquest is a biological vestige imposed by evolution. However, Bostrom's *Orthogonality Thesis* and *Instrumental Convergence Thesis* offer strong counters to this position.[xxxii] Orthogonality suggests that the final goals of any intelligent agent and intelligence itself exist on orthogonal axes, meaning that they are completely independent. In other words, any level of intelligence can be combined with any set of final goals. This dispenses with the idea that intelligent agents will converge to similar goals and that we could anticipate a second intelligent "species" with whom we share our planetary resources would ultimately have final goals like ours. Furthermore, instrumental convergence suggests that there exists a set of instrumental *sub-goals* that may be common among any set of intelligent agents if they increase the chance of that agent's *final goal* being realized. A non-exhaustive list of potential sub-goals that fit this definition include self-preservation, resource and power acquisition, technological development, and self-improvement.[xxxiii] Many of these goals are zero-sum games that could reasonably lead to interspecific conflict.

Another dire possibility would be the emergence of artificial superintelligence (ASI), a subset of AGI that exceeds human cognitive capabilities.[xxxiv] Achieving superintelligence would be a reasonable sub-goal for any intelligent agent given the instrumental sub-goals of self-improvement, resource and power acquisition, and technological development. In other words, any AI that *can* improve its own code would likely be motivated to do so. Given enough compute capacity and other resources, this could very quickly lead to an intelligence explosion from recursive self-improvement.[xxxv] In this scenario, humans would relinquish their position as the dominant intelligent species on the planet with potential calamitous consequences. Like the Neanderthals, our control over our future, and even our very existence, may end with the introduction of a more intelligent competitor.

It stands to reason that an out-of-control technology, especially one that is goal-directed like AI, would be a good candidate for the Great Filter. While there is currently no astrobiological evidence that this is the case, postbiological approaches to astrobiology are considered viable.[xxxvi,xxxvii] The discovery of artificial extraterrestrial intelligence without concurrent evidence of a pre-existing biological intelligence would certainly move the needle in favor of advanced AI being a cause of the Great Filter. Furthermore, if Bostrom's instrumental convergence thesis is correct, then any intelligent agent would likely develop sub-goals that are inherently competitive. Thus, we would not be anthropomorphizing to suggest that an extraterrestrial artificial intelligence would behave similarly to what we observe in their terrestrial counterparts, i.e., in ways that potentially outcompete its biological predecessors. Moreover, given the instrumental sub-goal of resource and power acquisition, actively signaling our existence in a way detectable to such an extraterrestrial AI may not be in our best interest. Any competitive extraterrestrial AI may be inclined to seek resources elsewhere – including Earth. If this is the case, perhaps the Great Filter is the competitive AI offspring from whatever biological extraterrestrial species developed it first, now moving through the cosmos extinguishing any biological competitors who make themselves known. While it may seem like science fiction, it is probable that an out-of-control, agential technology like AI would be a likely candidate for the Great Filter – whether organic to our planet, or of extraterrestrial origin. We must ask ourselves; how do we prepare for this possibility?

**Implications for the Intelligence Community**

    For answers to some problems, the Intelligence Community must learn to look to the stars.  Examining risk through the lens of the Great Filter is a way to mitigate the effects of anthropic bias, thus reducing the chance of underestimating global catastrophes.  Taking a cosmic view provides a unique perspective that is free from our default human-centric worldview.  It should also give us pause because it reveals that the greatest risks to humanity appear to be ones of our own making.  This humbling realization should inform how we prioritize long-term risk mitigation strategies, especially ones dealing with disruptive technologies like artificial intelligence.

    When assessing global catastrophic risks, we very often succumb to *non-extensional reasoning*, which is a bias that arises when we evaluate descriptions of events and not the events themselves.[xxxviii]  The logical *extension* of catastrophic risks is that you, your family, loved ones, and others whom you care about will die, perhaps violently, yet this outcome is not considered. Cataclysmic events like human extinction are often met with a shrug. We imagine science fiction novels, movies, and events far into the future, but not the actual effects of the event if it was to happen at this very moment.  The possibility of a child or partner having a terminal illness is a terrible thought because we focus on the inevitable outcome of the event, yet the outcome would be the same if the event was human extinction.  We anchor to what is familiar.  Terminal illness is a tragedy that we can imagine because it happens often.  The same can't be said for human extinction.  When it comes to out-of-control AI leading to catastrophe, we have nothing real to anchor to (only fiction like the *Terminator* movie series and the science fiction film *2001: A Space Odyssey*), so we don't focus on the outcome.  We dismiss the event as improbable when the reality may be quite different.

In light of this awareness, it would be prudent to consider a global moratorium on some forms of AI research. While a unitary effort like this is no trivial task, the potential consequences of not doing so could be dire. In fact, there is strong evidence that existential risk mitigation should be a global priority.[xxxix] The United States should lead the way in this effort, and the IC can contribute by working with our allies to analyze and understand the risks of AI from the perspective of the Great Filter. Generating a global conversation around AI safety is the first step, but it must begin soon given the rapid pace of AI development. As of this writing, the crowdsourced forecasting site *Metaculus* predicts that artificial general intelligence will become a reality sometime between 2026 and 2044, with a median prediction of 2032.[xl] Notably, the predicted date continues to trend toward the present. Even if AGI doesn't become a reality within the next decade, the subversive agential behavior of contemporary *weak* AI systems is extremely disconcerting. This problem will only become more severe as this technology advances. The Intelligence Community should prioritize getting in front of this problem now *before* AI becomes so advanced and ubiquitous that it can no longer be controlled. Once the problem is evident, it may be too late to intervene. It's not an exaggeration to suggest that, if we fail to do this, we could lose control of our own destiny. AI may very well be the last thing we ever invent – the Great Filter that extinguishes our future in a silent cosmos.[7]

--
Dr. Mark Bailey is Chair of the Cyber Intelligence and Data Science Department at National Intelligence University and Co-Director of the Data Science Intelligence Center. He is also a Distinguished Senior Fellow for Complexity, AI, and Risk at the Center for the Future Mind. Previously, he worked as a data scientist on several AI programs in the U.S. Department of Defense and the IC. He is also a Major in the U.S. Army Reserve.

---

[7] The author would like to thank Mark Heiligman, Susan Schneider and Mitch Simmons for their helpful comments.

[i] Jones, Eric M. *Where is Everybody? An Account of Fermi's Question*, U.S. Department of Energy, 1985, https://www.osti.gov/biblio/5746675.

[ii] The Drake Equation. *SETI Institute*. Accessed 5 April 2023. https://www.seti.org/drake-equation-index.

[iii] Hill, Michelle L., *et al*. "A Catalog of Habitable Zone Exoplanets" *The Astronomical Journal*. Vol. 165(34), 2023.

[iv] National Aeronautics and Space Administration. *Exploring Alien Worlds*. Accessed 6 April 2023, https://exoplanets.nasa.gov/alien-worlds/historic-timeline/#first-planetary-disk-observed.

[v] Jones, 1985.

[vi] Ward, Peter D., and Brownlee, Donald. *Rare Earth: Why Complex Life is Uncommon in the Universe*. New York, Springer, 2004.

[vii] Peacock, John A. *Cosmological Physics*. Cambridge, Cambridge University Press, 1998.

[viii] Hill, 2023.

[ix] Hanson, Robin. The Great Filter – Are Se Almost Past It? 1998, https://mason.gmu.edu/~rhanson/greatfilter.html.

[x] Cirkovic, Milan A. "Observation Selection Effects and Global Catastrophic Risk." In *Global Catastrophic Risks*. Ed. N. Bostrom and M. Cirkovic. Oxford, Oxford University Press, 2008.

[xi] Leslie, John. *The End of the World: The Science and Ethics of Human Extinction*. Routledge, NY, Taylor & Frances, 1996.

[xii] Bostrom, Nick. "The Doomsday Argument is Alive and Kicking." *Mind*. Vol. 108(431), 1999.

[xiii] Bostrom, Nick. *Anthropic Bias: Observation Selection Effects in Science and Philosophy*. New York, NY. Routledge, 2002.

[xiv] *Ibid.*

[xv] Leslie, 1996.

[xvi] Snyder-Beattie, Andrew E., *et al*. "An Upper Bound for the Background Rate of Human Extinction." *Nature Scientific Reports*. Vol. 9, 2019.

[xvii] Cohen, Irun R., and Harel, David. "Explaining a Complex Living System: Dynamics, Multi-Scaling and Emergence." *Journal of the Royal Society Interface*. Vol. 4(13), 2007.

[xviii] Jeffery, Kate, *et al*. "On the Statistical Mechanics of Life: Schroedinger Revisited." *Entropy*. Vol. 21, 2019.

[xix] Merino, Nancy, *et al*. "Living at the Extremes: Extremophiles and the Limits of Life in a Planetary Context." *Frontiers in Microbiology*. Vol. 10, 2019.

[xx] Snyder-Beattie, 2019.

[xxi] Hendrix, Cullen S. "Climate Change and Conflict." *Nature Reviews Earth & Environment*. Vol. 4, 2023.

[xxii] Haijun, Song, *et al*. "Thresholds of Temperature Change for Mass Extinctions." *Nature Communications*. Vol. 12, 2021.

[xxiii] "Who Owns the World's Nuclear Weapons?" *Federation of American Scientists.* Accessed 7 April 2023, https://fas.org/issues/nuclear-weapons/status-world-nuclear-forces/.

[xxiv] Vilone, Giulia and Longo, Luca. "Notions of Explainability and Evaluation Approaches for Explainable Artificial Intelligence." *Information Fusion*. Vol. 76, 2021.

[xxv] Lavoipierre, Ange. *A Journey Inside Our Unimaginable Future*. Australia Broadcasting Corporation. Accessed 7 April 2023, https://www.abc.net.au/news/2022-11-26/loab-age-of-artificial-intelligence-future/101678206.

[xxvi] "Pause Giant AI Experiments: An Open Letter." *Future of Life Institute*. 2023. https://futureoflife.org/open-letter/pause-giant-ai-experiments/.

[xxvii] Ngo, Richard. *AGI Safety from First Principles*. 2020.

[xxviii] OpenAI. *GPT-4 System Card*. 2023, https://cdn.openai.com/papers/gpt-4-system-card.pdf.

[xxix] Bailey, Mark M. *Understanding and Mitigating the Long-Term Risks of AI Operationalization*. NIU Press, 2023, https://ni-u.edu/wp/wp-content/uploads/2023/03/NIUShort_20230301_DNI_2023_00853.pdf.

[xxx] Ngo, 2020.

[xxxi] Ramirez Rozzi, Fernando V., *et al.* "Neandertal and modern Humans at Les Rois." *Journal of Anthropological Sciences*. Vol. 87, 2009.

[xxxii] Bostrom, Nick. "The Superintelligent Will: Motivation and Instrumental Rationality in Advanced Artificial Agents." *Minds and Machines*. Vol. 22(2), 2012.

[xxxiii] Ngo, 2020.

[xxxiv] Bostrom, Nick. *Superintelligence: Paths, Dangers, Strategies*. Oxford, Oxford University Press, 2014.